\documentclass[%
 aip,
 amsmath,amssymb,
 reprint,%
]{revtex4-1}
\usepackage{physics}

\usepackage{amsmath}
\usepackage{mathptmx,bm}
\usepackage{graphicx}
\usepackage{dcolumn}
\usepackage{tabularx}
\usepackage{epsf}
\usepackage{color}

\usepackage{hyperref}
\hypersetup{breaklinks,colorlinks,linkcolor=blue,citecolor=blue,urlcolor=blue}

\usepackage{epstopdf}%
\usepackage{verbatim}
\usepackage[english]{babel}
\newcommand{\bea}{\begin{eqnarray}}
\newcommand{\eea}{\end{eqnarray}}
\usepackage{amsfonts}

\usepackage{graphicx}
\usepackage{dcolumn}
\usepackage{bm}

\usepackage[utf8]{inputenc}
\usepackage[T1]{fontenc}
\usepackage{mathptmx}
\usepackage{etoolbox}

\makeatletter
\def\@email#1#2{%
 \endgroup
 \patchcmd{\titleblock@produce}
  {\frontmatter@RRAPformat}
  {\frontmatter@RRAPformat{\produce@RRAP{*#1\href{mailto:#2}{#2}}}\frontmatter@RRAPformat}
  {}{}
}%
\makeatother

\begin{document}


\title{Quantum Thermal Transport Beyond Second Order with the Reaction Coordinate Mapping}

\author{Nicholas Anto-Sztrikacs}
\affiliation{Department of Physics, 60 Saint George St., University of Toronto, Toronto, Ontario, Canada M5S 1A7}

\author{Felix Ivander}
\affiliation{Chemical Physics Theory Group, Department of Chemistry and Centre for Quantum Information and Quantum Control,
University of Toronto, 80 Saint George St., Toronto, Ontario, M5S 3H6, Canada}

\author{Dvira Segal}
\affiliation{Chemical Physics Theory Group, Department of Chemistry and Centre for Quantum Information and Quantum Control,
University of Toronto, 80 Saint George St., Toronto, Ontario, M5S 3H6, Canada}
\affiliation{Department of Physics, 60 Saint George St., University of Toronto, Toronto, Ontario, Canada M5S 1A7}
\email{dvira.segal@utoronto.ca}

\date{\today}

\begin{abstract}
Standard quantum master equation techniques 
such as the Redfield or Lindblad equations are perturbative to second order in the microscopic system-reservoir coupling parameter 
$\lambda$. 
As a result, characteristics of dissipative systems, 
which are beyond second order in $\lambda$, are not captured by such tools. 
Moreover, if the leading order in the studied effect 
is higher-than-quadratic in $\lambda$, a second-order description 
fundamentally fails even at weak coupling.
Here, using the reaction coordinate (RC) quantum master equation  framework, we are able to 
investigate and classify higher-than-second order transport mechanisms.  
This technique, which relies on the redefinition of the system-environment boundary, 
allows for the effects of system-bath coupling to be included to high orders. 
We study steady-state heat current beyond second-order in two models: 
The generalized spin-boson model with non-commuting system-bath operators and a three-level ladder system. 
In the latter model heat enters in one transition and it is extracted from a different one. 
Crucially, we identify two transport pathways:
(i) System's current, where heat conduction  is mediated by transitions in the system, 
with the heat current scaling as $j_q \propto \lambda^2$ to lowest order in $\lambda$. 
(ii) Inter-bath current,
with the thermal baths directly exchanging energy between them, facilitated by the bridging quantum system.
To the lowest order in $\lambda$, this current scales as $j_q \propto \lambda^4$. 
These mechanisms are uncovered and examined using  numerical and analytical tools.
We contend that the RC mapping brings already at the level of the mapped Hamiltonian much insights on transport characteristics.
\end{abstract}

\maketitle

\section{Introduction}
\label{Sec:Introduction}

Dissipative quantum impurity systems \cite{Weiss} constitute the core of 
quantum thermal junctions and machines. Moreover, dissipative systems are central to problems of charge transport at the nanoscale, chemical reactions in the condensed phases \cite{Nitzan}, and light-matter interaction effects in quantum optics \cite{Breuer}. More recently, dissipative systems have received substantial attention in the field of quantum thermodynamics where one is interested in understanding---and drawing on---quantum mechanical effects for the design of thermal machines \cite{Anders_2016,Goold_2016,Kosloff_2013}.

Perturbative quantum master equation (QME) techniques are commonly used for studying quantum dissipative systems due to their simplicity and potential for analytical insights \cite{Breuer,Anders}. 
In particular, the popular Redfield approach relies on the Born approximation (as well as other assumptions), where the system-environment coupling strength $\lambda$ is assumed weak. In such methods, the dynamics and steady state behavior of the dissipative system is obtained to second order in the microscopic system-environment coupling parameter, with the heat current scaling as $j_q\propto \lambda^2$, see \cite{SegalQME,comment2}.

This outlined perturbative approach is valid and accurate in the weak coupling regime---as long as
the second order expansion describes the leading-order physics. 
However, there exist systems for which the first non-trivial term in the expansion (say for the heat current) 
goes beyond second order in the system-reservoir coupling parameter. 
In such cases, a second order QME approach completely fails, and it offers incorrect predictions
even in the asymptotically weak-coupling regime where QMEs are typically expected to thrive. 
To handle such situations, more accurate techniques, which capture the system-reservoir coupling beyond second order
are required. 
We stress that this aspect is distinct from the commonly addressed issue of 
inaccurate predictions due to strong system-bath couplings. Instead, here we are interested in systems 
{\it in the asymptotically weak system-bath coupling regime}, yet where standard QMEs completely fail since
the lowest order term in the behavior of the heat current goes beyond second-order in $\lambda$.

The reaction coordinate mapping method is well-suited to describe quantum dissipative systems beyond second order \cite{Garg,Ank04,Burghardt1,Burghardt2,Thoss17,Nazir18}. This method redefines the system-environment boundary with the identification and extraction of a central (collective) degree of freedom of the environment, termed the reaction coordinate (RC). 
The quantum system is then expanded to become an “extended system”, 
which comprises the original, pre-mapped quantum system, 
the RC, and the interaction of the RC with the original system. 
With the inclusion of the RC within the system itself, 
the microscopic coupling parameter ($\lambda$) of the original system to the surroundings is now 
included to high orders within the reaction coordinate framework. 
As such, the RC mapping should be a viable tool to explore high-order transport phenomena. 

The advantage of the exact RC mapping becomes clear
once adopting a second-order QME technique, such as the Born-Markov Redfield (BMR) equation, 
to simulate the dissipative behavior of the {\it extended} system. 
The resulting technique, termed the RC-QME method, is nonperturbative in the original coupling parameter $\lambda$. 
Since the method enables cheap computations, 
it has been used in studies involving strongly-coupled system-reservoirs. 
For example, it was utilized for studying quantum dynamics and steady state behavior of impurity models 
\cite{NazirPRA14,Nazir16,Camille}, thermal transport in nanojunctions \cite{Correa19,Nick2021}, the operation of quantum thermal machines \cite{Newman_2017, Newman_2020,Strasberg_2016,QAR-Felix}, 
transport in electronic systems \cite{GernotF,GernotF2,McConnel_2021}. 
Besides capturing system-bath coupling to high orders,  the RC-QME method reports on non-Markovian effects \cite{Nickdecoh}. This is because correlations building up between the system and reservoirs 
are maintained throughout the evolution of the extended system. 

The RC method was recently implemented for studying 
the nonequilibrium spin-boson model (NESB) at strong coupling \cite{Nick2021}, 
a quintessential model in the quantum thermodynamics field. 
The model includes an impurity two-level system (spin) interacting with two thermal harmonic-oscillator
 environments held at different temperatures, allowing for a steady state heat current to flow from one 
reservoir to the other. The model has been recently  realized experimentally using superconducting quantum circuits demonstrating a thermal diode effect \cite{PekolaE}. Furthermore, the NESB model has been studied extensively using  
perturbative and numerically-exact techniques;
a partial list includes Refs. 
\cite{PRL05,Thoss08, Thoss10,Nicolin11, Saito13, Segal14, Cao15,CaoNEPTRE,Tanimura16, Qi17,Bijay17,Liu17, Kilgour19,Kelly19,Gabe18,Gabe19,Gabe21,Wu20,Aurell20,Nick2021}. 

The spin-boson model has been recently extended to explore dynamics and transport effects 
using a general coupling operator with both diagonal and off-diagonal couplings \cite{Camille,
Gernot16, Gernot17, Zhao16,Nalbach18,Nalbach19, Fingerhut20,Archak, Junjie20,Belyansky2021, Zhang2021,Chen2022}. 
For heat transport, the coupling operators to the different reservoirs do not commute.
As a result, the model manifests the emergence of two distinct transport regimes \cite{Junjie20}: 
(i) The current flows sequentially from the hot to the cold reservoir mediated by the excitation of the central spin.
 (ii) Heat flows directly from the hot to the cold reservoir, 
apparently without exciting the central spin, 
in a manner reminiscent of the tunneling-superexchange behavior \cite{Scholes03,Segal11}.
This direct bath-to-bath current between reservoirs in the generalized, non-commutative spin-boson model 
is an example of a transport phenomenon that emerges 
beyond second order in perturbation theory in the system-bath coupling energy.

In this paper,
 we use the RC-QME method to probe beyond-second-order effects in nonequilibrium 
steady-state.  We analyze two impurity models
in which the heat current scales as $j_q\propto\lambda^4$
 even in the asymptotic weak coupling limit, $\lambda\to 0$, 
and provide analytical intuition into the unusual transport phenomena observed. 

This work contributes the following:
(i) Applying the RC mapping we immediately bring---at the level of mapped Hamiltonian and without performing transport calculations---fundamental understanding of different thermal transport mechanisms.
(ii) We establish the RC-QME method for involved nonequilibrium quantum transport applications 
by showing that it realizes both system's and direct-bath currents. 
(iii) We bring forward a new, ladder model for heat transport, with heat inputted from one step (transition) 
and outputted through a different transition.

The paper is organized as follows. 
In Section \ref{Sec:Currents}, we formally discuss two contributions to the heat current, the
system's current and the inter-bath current. 
We analyze the generalized spin-boson model in Sec. \ref{Sec:SB-Model} using the RC mapping
based on analytical considerations and numerical simulations.
In Sec. \ref{Sec:ladder}, we study a three-level ladder model for heat transport, which relies on an inter-bath transport pathway at weak coupling.
We conclude in Section \ref{Sec:Conclusion}.   


\section{Classification of heat currents} 
\label{Sec:Currents}

Model Hamiltonians for quantum dissipative systems are typically partitioned into contributions from the system, 
its surroundings, and the interaction between them. 
For a system coupled to two independent reservoirs, the total Hamiltonian may be represented as follows,
\bea
\hat{H} = \hat{H}_S + \kappa_R\hat{H}_{I,R} + \kappa_L\hat{H}_{I,L} + \hat{H}_{B,R} + \hat{H}_{B,L},
\eea
where $\kappa_{R,L}$ are dimensionless parameters used for bookkeeping the coupling between the system and the environments. 
Without invoking assumptions on the microscopic form of the model, a formal expression for the average energy (heat, in the present case) current, flowing out of one of the reservoirs may be obtained from Heisenberg's equation of motion. For example, the average heat current flowing out of the left reservoir is simply
(we work in units of $\hbar\equiv 1$ and $k_B\equiv 1$)
\bea
j_q\equiv -   \langle \dot{\hat{H}}_{B,L} \rangle = -i\langle [\hat{H} , \hat{H}_{B,L}] \rangle,
\eea
where the dot represents a time derivative. 
Utilizing the equation of motion for $\hat{H}_{I,L}$, and the fact that in steady-state
$\langle \dot{\hat{H}}_{I,L}\rangle =i\langle [\hat H,\hat H_{I,L}] \rangle=0$, 
one obtains a formal expression for the heat current as \cite{TanimuraBook} 
\bea
    j_q =  -i\kappa_L\langle [\hat{H}_S,\hat{H}_{I,L}] \rangle -i\kappa_L\kappa_R\langle [\hat{H}_{I,R},\hat{H}_{I,L}] \rangle.
    \label{eq:current}
\eea
The above expression contains two contributions to the steady state heat current, both stemming from the non-commutativity of two components of the total Hamiltonian. 
%
The first contribution arises due to the non-commutativity of the system Hamiltonian with the interaction Hamiltonian of the left reservoir, and is proportional only to the coupling parameter to the left reservoir, $\kappa_L$. Motivated by this, one defines this contribution as the {\it system current}. The second term originates from the non-commutativity of both interaction Hamiltonians, and it depends on {\it both} coupling parameters, $\kappa_L$ and $\kappa_R$. This term constitutes a form of higher order transport, which is defined as the {\it inter-bath current} \cite{Tanimura16}. 

Using the different dependencies on the coupling parameters in Eq. (\ref{eq:current}) as a guide, 
it was pointed out in Ref. \cite{TanimuraBook} that one can distinguish between two transport mechanisms in numerical simulations of the heat current, by observing how they scale with the coupling parameters. It is worthy of note, however, that standard QME techniques capture the effects of system-reservoir interactions only to the lowest order in $\kappa_{L,R}$. 
As a result, the inter-bath current, which is higher order in the coupling parameters, cannot be captured by second-order QME approaches.

\begin{figure}
 \centering
\includegraphics[width=0.8\columnwidth]{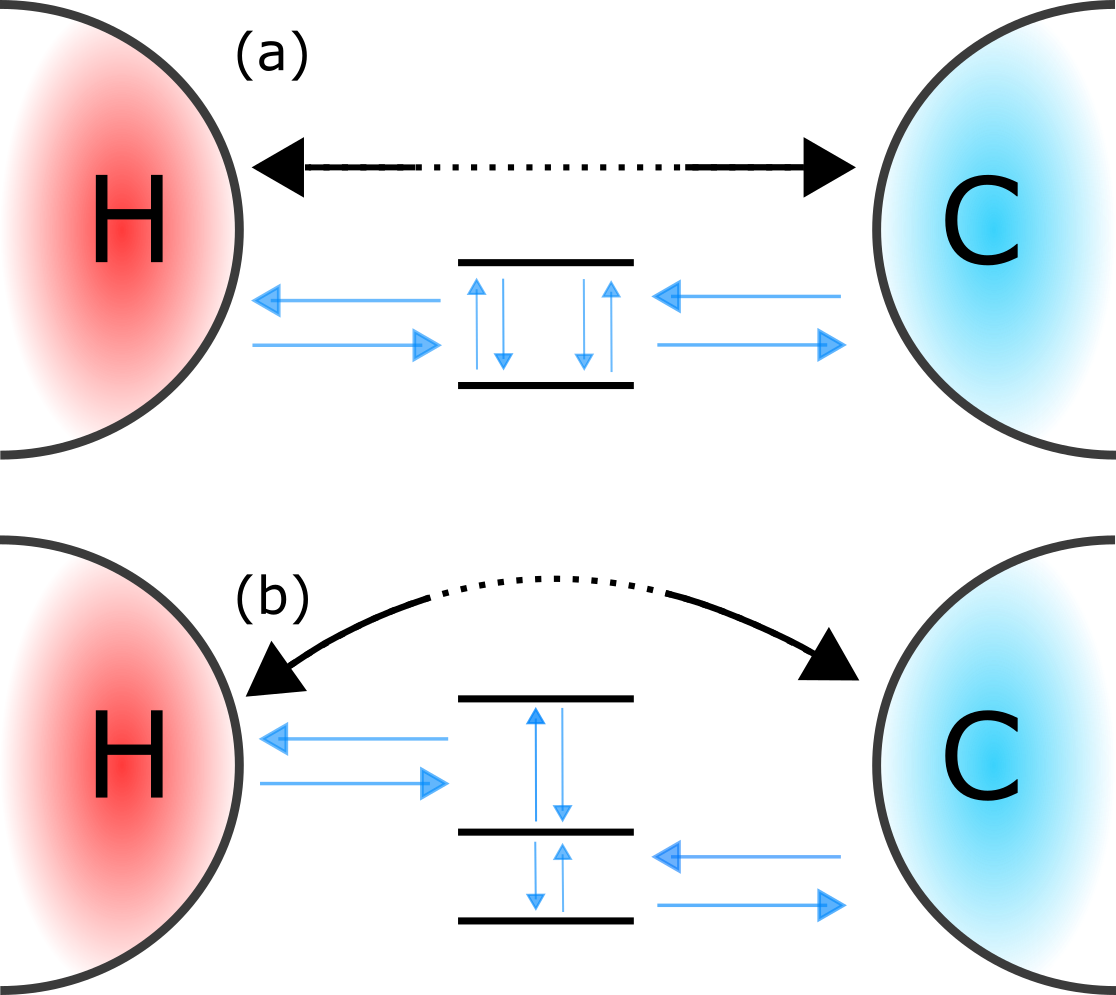} 
\caption{Schemes of heat transport models investigated in this work: 
(a) The generalized NESB model, and the (b) ladder model.
We mark with straight arrows heat exchange processes that involve the system, contributing to the system's current. 
The dashed-straight arrows illustrate the inter-bath heat current, of direct bath-to-bath energy exchange using the system as a bridge.
    }
    \label{fig:Diagram}
\end{figure}

With this in mind, in order to probe both transport mechanisms,
in this work we focus on models where one term in Eq. (\ref{eq:current}) dominates over the other.
%
In what follows, we study two models of quantum heat transport:
(i) the generalized spin-boson model
and (ii) a three-level ladder model.
In both cases, we focus on the regime of weak system-bath couplings,
when only lowest order terms in the expressions  Eq. (\ref{eq:current}) contribute.
The first model, the generalized spin-boson model, shows rich behavior:
by tuning the coupling operators, the relative dominance of the different contributions in Eq. (\ref{eq:current}) is controlled.
In the ladder model, which we construct here, the system's current does not contribute to  
heat flow at weak couplings.
Both models and the different currents are illustrated in Figure \ref{fig:Diagram}.

\section{The generalized spin-boson model}
\label{Sec:SB-Model}

\subsection{Hamiltonian and the reaction coordinate mapping}

Our first model is the generalized spin-boson model as introduced in 
Ref. \cite{Junjie20}. 
The model consists of a single spin interacting with two bosonic environments maintained at different temperatures. 
The nontrivial aspect of the generalized model, compared to the ``standard" one, is that system's operators that are coupled to the baths do not commute with each other; the left (hot) bath is coupled to the system via $\hat{\sigma}_x$ while the right (cold) reservoir is coupled with $\hat{\sigma}_{\theta} = \hat{\sigma}_z \cos(\theta) + \hat{\sigma}_x \sin(\theta)$ with $0 \leq \theta \leq \pi/2$  
denoting the non-commutativity parameter. This model is expressed as
\bea
    \nonumber
    \hat{H}_{SB} = &&\frac{\Delta}{2}\hat{\sigma}_z + \hat{\sigma}_x \sum_k f_{k,L} (\hat{c}_{k,L}^{\dagger} + \hat{c}_{k,L}) 
    \\
    &+& \hat{\sigma}_{\theta} \sum_k f_{k,R} (\hat{c}_{k,R}^{\dagger} + \hat{c}_{k,R}) 
    + \sum_{k,\alpha \in \{R,L\}} \nu_{k,\alpha} \hat{c}_{k,\alpha}^{\dagger} \hat{c}_{k,\alpha}.
    \label{eq:HSB}
\eea
The system includes a spin with energy splitting $\Delta$; $\hat{\sigma}_{x,y,z}$ are the Pauli matrices. $\hat{c}_{k,\alpha}^{\dagger}$ ($\hat{c}_{k,\alpha}$) are the creation (annihilation) operators of the two bosonic environments enumerated by $\alpha=L,R$. 
$f_{k,\alpha}$ denotes the coupling energies between the system and the $k$th mode of the $\alpha$ environment of frequency $\nu_{k,\alpha}$. The coupling of the spin to the baths is captured by the spectral density function $J_{SB,\alpha}(\omega) = \sum_k f_{k,\alpha}^2 \delta(\omega - \nu_{k,\alpha})$. The setup is depicted in Figure \ref{fig:Diagram} (a). 
For reasons that will soon become evident, we choose the spectral density function of the environments to be a structured, Brownian function
\bea
    J_{SB,\alpha}(\omega) = \frac{4\gamma_\alpha \omega \Omega_\alpha^2 \lambda_\alpha^2}{(\omega^2 - \Omega_\alpha^2)^2 + (2\pi \gamma_\alpha \Omega_\alpha \omega)^2}.
    \label{eq:JBrownian}
\eea
Here, $\lambda_\alpha$ denotes the system-environment coupling
strength, $\gamma_\alpha$ is a measure for the width of the spectral density function, and $\Omega_\alpha$ is the frequency about which the spectral density function is peaked.

The microscopic system-bath coupling parameters are $f_{k,\alpha}$, 
which are the interaction energies between
the system and the $k$th mode of the $\alpha$ bath.
The spectral density function weights these coupling strengths squared with the density of states. In the Brownian model Eq. (\ref{eq:JBrownian}),
 $J_{SB,\alpha}(\omega)\propto \lambda_{\alpha}^2$, with $\lambda_{\alpha}$ taking the role of system-bath coupling parameter now that the bath is assumed continuous and of a certain density of states.
In what follows, we use $\lambda_{\alpha}$ as a measure for the coupling energy.

To go beyond a standard second-order QME treatment---and uncover higher order transport phenomena---we implement the reaction coordinate QME technique. This method builds on an exact Hamiltonian mapping where a collective coordinate (reaction coordinate) is extracted from each reservoir and included as part of the system's Hamiltonian. This mapping thus redefines the system-environment boundary. Performing a reaction coordinate mapping on both baths leads to the reaction coordinate generalized spin-boson (SB-RC) Hamiltonian,
\bea
    \nonumber
    \hat{H}_{SB-RC} &= &\frac{\Delta}{2} \hat{\sigma}_z + \Omega_L\hat{a}_L^{\dagger}\hat{a}_L + \Omega_R\hat{a}_R^{\dagger}\hat{a}_R + \lambda_L \hat{\sigma}_x(\hat{a}_L^{\dagger} + \hat{a}_L)
    \\ \nonumber
    &+& \lambda_R \hat{\sigma}_{\theta} (\hat{a}_R^{\dagger} + \hat{a}_R) + \sum_{k,\alpha} \frac{g_{k,\alpha}^2}{\omega_{k,\alpha}} (\hat{a}_\alpha^{\dagger} + \hat{a}_\alpha)^2
     \nonumber\\
    &+& \sum_{\alpha} (\hat{a}_\alpha^{\dagger} + \hat{a}_\alpha) \sum_k g_{k,\alpha} (\hat{b}_{k,\alpha}^{\dagger} + \hat{b}_{k,\alpha})  
    + \sum_{k,\alpha} \omega_{k,\alpha}  \hat{b}_{k,\alpha}^{\dagger} \hat{b}_{k,\alpha}.
   \nonumber\\
\label{eq:HSBRC}
\eea
In this expression, $\hat{a}_{\alpha}^{\dagger}$ ($\hat{a}_{\alpha}$)  and $\hat{b}_{k,\alpha}^{\dagger}$ ($\hat{b}_{k,\alpha}$) represent the creation (annihilation) operators of the reaction coordinates and the residual baths, with frequencies $\Omega_{\alpha}$ and $\omega_{k,\alpha}$, respectively. The reaction coordinates couple to the system via coupling parameter $\lambda_{\alpha}$. In turn, the residual baths couple to the reaction coordinates via a coupling $g_{k,\alpha}$, which is captured via a spectral density function 
$J_{SB-RC,\alpha}(\omega) = \sum_k g_{k,\alpha}^2 \delta(\omega - \omega_{k,\alpha})$.    
It can be shown that if the spectral density function of the original, pre-mapped reservoir is of Brownian form, Eq. (\ref{eq:JBrownian}), then the spectral density function of the residual bath becomes Ohmic with an 
infinite high frequency cutoff after the reaction coordinate mapping \cite{NazirPRA14,Nick2021},
\bea
  J_{SB-RC,\alpha}(\omega)=\gamma_{\alpha}\omega e^{-\abs{\omega}/\Lambda_{\alpha}}.
  \label{eq:JOhmic}
\eea
In this representation, $\gamma_{\alpha}$ is a dimensionless coupling constant between 
the $\alpha$th reaction coordinate and its residual bath, and
$\Lambda_{\alpha}$ is the cutoff frequency of the $\alpha$th bath. 
The first two lines of the SB-RC Hamiltonian, Eq. (\ref{eq:HSBRC}), represent the {\it extended system}, comprising of the spin, the two RC harmonic oscillators, and their interactions. Note the quadratic term in the RC, which emerges after the mapping. 
The last line contains the residual baths and the interaction between the reaction coordinates and the residual baths. 

\subsection{Transport mechanisms based on the RC mapping}
\label{Sec:mech}

Equation (\ref{eq:current}) suggests two distinct contributions to the heat transfer, which may be formally understood as arising from the non-commutativity of components of the total Hamiltonian. However, this formally-exact expression does not provide fundamental understanding of the mechanisms behind different transport pathways.
The standard way to approach this problem and gain deeper understanding 
is to preform numerical simulations of the heat current.
This task is typically nontrivial: It requires solving quantum equations 
of motion in steady state, and calculating expectation values of different observables.
Nonetheless, the advantage of the RC mapping is
that already at the level of the mapped Hamiltonian, Eq. (\ref{eq:HSBRC}), and without performing transport calculations,
we can gain substantial insights onto transport behavior in the model.

In what follows, we examine contributions to the heat current, both
the system's current and the inter-bath currents, by studying the
SB-RC Hamiltonian, Eq. (\ref{eq:HSBRC}).
First, we use numerical tools and study the eigenspectrum and the coupling pattern of the SB-RC Hamiltonian.
We complement this numerical investigation with an analytical analysis: We apply the small-polaron transformation onto the SB-RC Hamiltonian, and interpret the contribution of different terms to the heat current.
%

\subsubsection{Numerical diagonalization}

We diagonalize the extended system [first five terms in Eq. (\ref{eq:HSBRC})] 
and denote the eigenenergies by $E_n$,
$\hat H_S^D=\sum_n E_n |n\rangle \langle n|$
with $n$ the eigenvalue index.
We further write down the system-bath coupling 
(the first term in the last line of Eq. (\ref{eq:HSBRC})) in this basis as
\bea
\hat H_{I,\alpha}^D =\left[\sum_{m,n}(\hat{S}_{\alpha}^D)_{m,n}|m\rangle \langle n|\right]
\left[\sum_{k}g_{k,\alpha}\left(\hat b_{k,\alpha}^{\dagger} + \hat b_{k,\alpha}\right) \right]. 
\label{eq:HID}
\eea
The coefficients $(\hat{S}_{\alpha}^D)_{m,n}$ are dictated by the original 
form of couplings and the diagonalization of 
$(\hat a_{\alpha} + \hat a_{\alpha}^{\dagger})$.

\begin{figure}[hbt]
    \centering
    \includegraphics[width=\columnwidth]{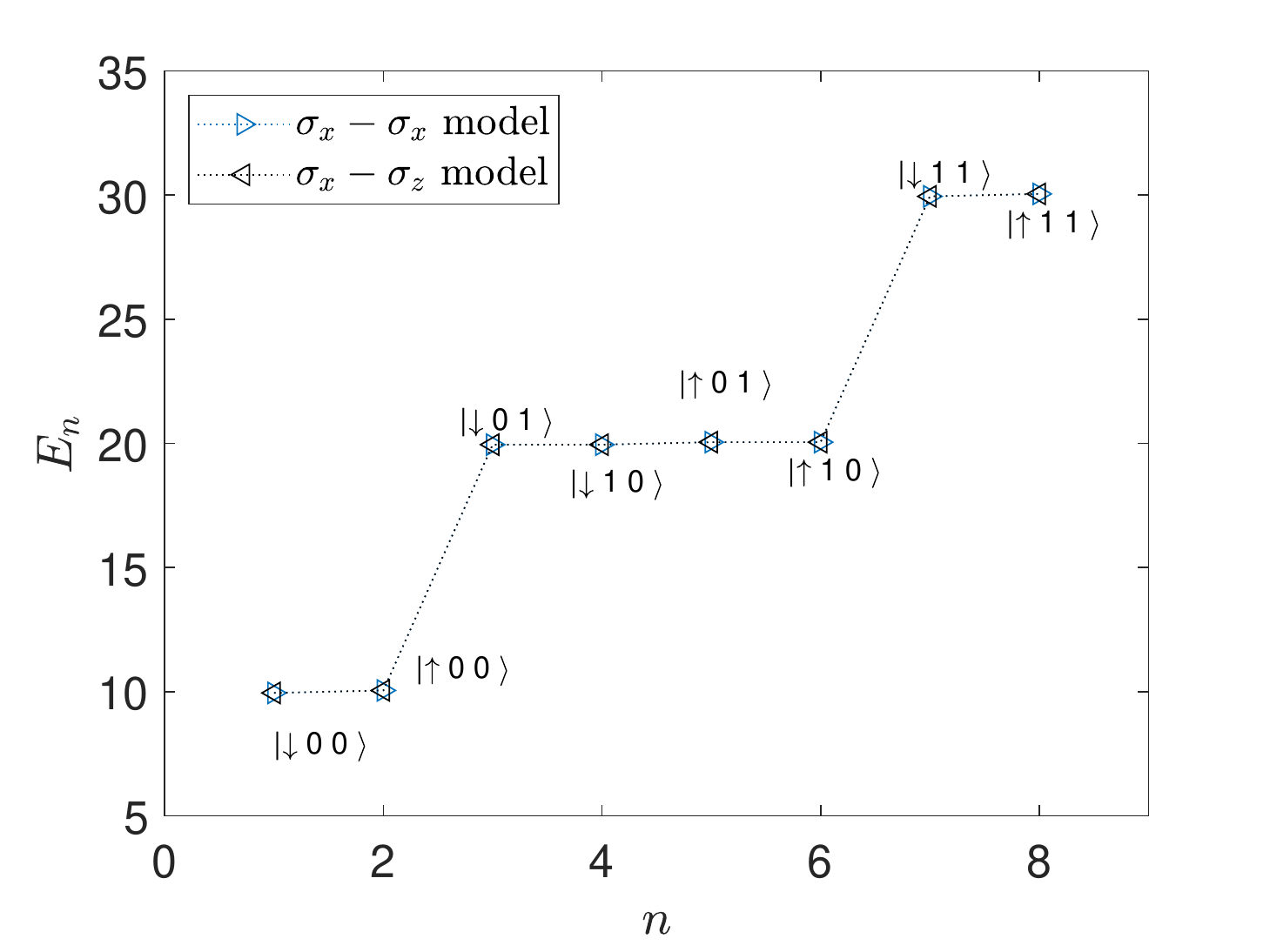} 
    \caption{Eigenspectrum of the extended system Hamiltonian Eq. (\ref{eq:HSBRC}) as a function of eigenstate index $n$ for two non-commutativity angles: 
(i) The $\sigma_x - \sigma_x$ model with $\theta = \pi/2$.
(ii) The $\sigma_x - \sigma_z$ model with $\theta = 0$. 
Parameters are set to $\Delta = 0.1$, $\lambda_L = \lambda_R = \lambda = 0.01$, $\Omega_L = \Omega_R = 10$. 
Since $\lambda/\Delta \ll1$, we can interpret the eigenstates using the local-site basis representation, where levels are labeled according to the state of the spin and the occupation number of the reaction coordinates, $\ket{s,n_R,n_L}$.}
    \label{fig:Spectrum}
\end{figure}

Figure \ref{fig:Spectrum} displays the eigenspectrum $E_n$
as a function of the index $n$ for two angles:
(i) The choice $\theta = \pi/2$
corresponds to the standard spin-boson model, with both baths coupled to the system via $\sigma_x$.
This is the $\sigma_x-\sigma_x$ model.
(ii) With the angle $\theta = 0$, the left bath is coupled via $\sigma_x$ to the system, while the right bath couples
with a so-called diagonal form, to $\sigma_z$. This case is labeled as the $\sigma_x-\sigma_z$ model. 

The coupling parameters to the reaction coordinates, $\lambda_{\alpha}$, are kept small relative to the internal energy scale $\Delta$ to maintain an approximate local-basis representation, where states are labeled according to their spin state, $s$, and the reaction coordinate excitation numbers, $n_R$, and $n_L$,  combined into a single ket vector as  $\ket{s,n_R,n_L}$. 
Working in a small $\lambda_{\alpha}$ regime explains the significant overlap of eigenvalues of the $\sigma_x-\sigma_x$ and $\sigma_x-\sigma_z$ models, as the interaction $\lambda_{\alpha}$ is too weak to discern between the two models. 
We emphasize that maintaining the coupling parameters $\lambda_{\alpha}$ small
allows us to distinguish between the system's and the inter-bath currents,
see Eq. (\ref{eq:current}), as each term distinctly control the heat current.  

For simplicity, the energy spectrum displayed in Figure \ref{fig:Spectrum} was calculated 
after truncating the  harmonic manifold of the reaction coordinates, including only $M=2$ levels. 
However, the discussion holds analogously for larger $M$ values. 
The symmetry presented in the spectrum arises due to the values of the RC frequency $\Omega_{\alpha}$ employed, being equal for both reservoirs. Again, however, this is not an essential requirement, and asymmetric choices could be adopted as well.

The key observation from Fig. \ref{fig:Spectrum} is that there are three clusters of states, 
separated by energies of order $\Omega_{\alpha}$ (which is the frequency of the RCs).
The lowest two levels correspond approximately to RCs occupying their ground states while the two level system (spin) is either in its up or down state, $|\downarrow00\rangle$ and $|\uparrow00\rangle$.
From the other end, the highest two levels roughly correspond to the spin occupying its up or down states while the RCs now both occupy their first excited states, $|\downarrow 1 1 \rangle$ and $|\uparrow 1 1 \rangle$. 
The intermediate four levels approximately correspond to having one RC in its first excited state, and one remaining in its ground state, while again the spin can be up or down for both situations, e.g., a state of the form
$|\downarrow 1 0 \rangle$. 
\begin{figure*}[htb]
    \centering
    \includegraphics[width=1.6\columnwidth]{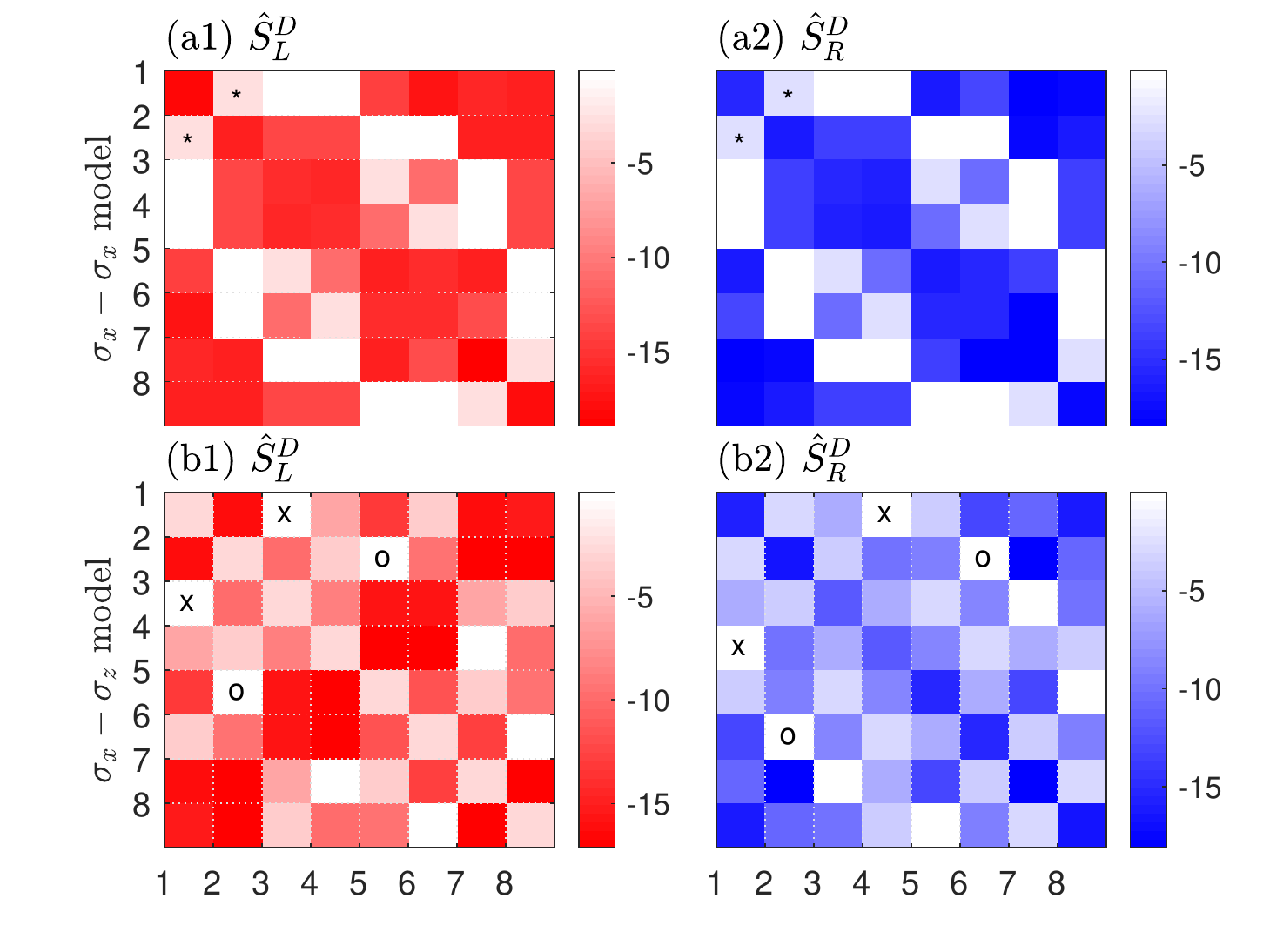} 
\caption{Heat map displaying the logarithm of the absolute value of the extended system's coupling operator 
$\log_{10}|\bra{n} \hat{S}_{L/R}^D \ket{m}|$, see Eq. (\ref{eq:HID}), 
with light color identifying large couplings.
Red (blue) panels show the coupling elements to the hot (cold) bath. 
The eigenstates enumerated by $n=1,2,...8$, are approximately local (spin and RCs) since $\lambda$ is small; the local basis closely coincides with the energy basis, as written in Figure \ref{fig:Spectrum}.
We mark by symbols (*,x,o) dominant processes.
(a1)-(a2)  $\sigma_x - \sigma_x$ model. Here, heat transfer
involves the excitation and de excitation of the system's spin, with
the dominant process (*)
$\ket{\downarrow,0,0} \leftrightarrow \ket{\uparrow,0,0}$.
  (b1)-(b2)  $\sigma_x - \sigma_z$ model.
Here, the RCs can exchange energy using the spin as a bridge: 
The dominant transitions are $\ket{\downarrow,0,0} \leftrightarrow \ket{\downarrow,0,1}$   and 
  $\ket{\downarrow,0,0} \leftrightarrow \ket{\downarrow,1,0}$
  (marked by x in (b1) and (b2), respectively) and similar transitions involving the spin up state $\ket{\uparrow,0,0} \leftrightarrow \ket{\uparrow,0,1}$   and 
  $\ket{\uparrow,0,0} \leftrightarrow \ket{\uparrow,1,0}$ (marked by o in (b1) and (b2), respectively).
  Parameters are identical to those used in Figure \ref{fig:Spectrum}.}
    \label{fig:Map}
\end{figure*}

The energy spectrum displays almost identical characteristics for the $\sigma_x-\sigma_x$ and the $\sigma_x-\sigma_z$ model. Their crucial distinction however 
shows up in the system-bath coupling pattern at the two ends.
Figure \ref{fig:Map} depicts the  matrix elements $(\hat{S}_{\alpha}^D)_{m,n}$ of the system's operators that are coupled to the baths, see Eq. (\ref{eq:HID}). Here, the $D$ superscript identifies that $|n\rangle$ and  $|m\rangle$ are eigenvectors of the extended system.
We display the elements coupled to the left (hot) and right (cold) baths and consider the angles $\theta = \pi/2$ (top) and $\theta = 0$ (bottom). We represent the matrix elements on a logarithmic scale, where larger values, and hence, more likely transitions are given by a lighter color. The symbols shown on the grid elements (*,x,o) mark transitions that contribute most significantly to the heat current. 

Figure \ref{fig:Map} (a1) and (a2) correspond to the $\sigma_x-\sigma_x$ model ($\theta = \pi/2$). These two panels show identical results given the symmetry in the coupling scheme. Since the coupling operators ($\sigma_x$) commute, the second term in Eq. (\ref{eq:current}) is  absent.
%
%
Assuming that temperatures are lower than the baths characteristic frequencies, $T_{\alpha}< \Omega_\alpha$
(though the temperature could be comparable to $\Delta$),
the dominant transitions are those with the RCs occupying their ground states and the baths creating excitations between the up and down spin states, $\ket{\downarrow 0 0} \leftrightarrow  \ket{\uparrow 0 0}$.
In Figure \ref{fig:Map}, we identify these transitions  with asterisks (*).

We now make our first observation: 
To low order in $\lambda$, the system's heat current, the first term in Eq. (\ref{eq:current}), involves the baths exchanging energy with the spin, or more generally, the central system. 
As a result, in heat current simulations at weak coupling we expect the system's current to be accurately predicted by low-order QME techniques.  

Figure \ref{fig:Map} (b1) and (b2) picture the coupling pattern for the $\sigma_x-\sigma_z$ model ($\theta = 0$) with $\hat{\sigma}_z$ coupled to the right bath.
The spatial symmetry in the spin-boson model is now broken.
The first term in Eq. (\ref{eq:current}), which is the system's current, vanishes, and only inter-bath current terms contribute towards the heat current. 
This is demonstrated in Figure \ref{fig:Map} (b1). 
For example the (x) pathway corresponds to the left (hot) bath allowing 
transitions between the approximate states 
$\ket{\downarrow 0  0} \leftrightarrow   \ket{\downarrow 0 1}$, while 
the right bath drives the transition $\ket{\downarrow 0  0} \leftrightarrow   \ket{\downarrow  1 0 }$. 
As a whole, the two baths exchange energy through the RCs, rather than the system.
Similarly,
(o) signifies transitions between $\ket{\uparrow 0 0} \leftrightarrow  \ket{\uparrow 0 1}$
and $\ket{\uparrow 0 0} \leftrightarrow  \ket{\uparrow  1 0 }$, again an inter-bath current.
%

Concerning the $\sigma_x-\sigma_z$ model, our second observation is that the second term in Eq. 
(\ref{eq:current}), the inter-bath current, 
describes coupled transitions of the RCs. It is therefore a higher order transport pathway corresponding to transitions in the baths; after all, the reaction coordinates are reservoirs' degrees of freedom. 
%

To summarize our analysis thus far, we  diagonalized numerically the extended Hamiltonian after
the RC mapping. Given the connectivity nature of the model,
even without performing transport calculations we are able to identify transport mechanisms and describe their characteristics:
The system's current involves spin transitions, and it will thus 
strongly depend on  the spin splitting, $\Delta$.
In contrast, in the $\sigma_x-\sigma_z$ model the spin plays an intermediating role, thus
(i) we expect the current to be largely independent of $\Delta$, but (ii) strongly
dependent on the RC frequency $\Omega$, or in other words, the spectral properties of the bath.

%

\subsubsection{Analysis in the polaron frame}
Complementing the numerical study, we include here a short analysis
that exposes the different transport contributions (system's current and inter-bath currents). This is achieved by performing
 an additional transformation on the SB-RC Hamiltonian Eq. (\ref{eq:HSBRC}).
%

Polaron theory is used to describe electron-phonon coupling effects by applying a unitary transformation onto the Hamiltonian. This results in the dressing of electronic degrees of freedom by the phonon degrees of freedom, forming an electron-phonon cloud (or a dressed electron, a quasiparticle). 
In the new basis, perturbative master equations in the tunneling splitting can be employed while still preserving aspects of strong system-reservoir  couplings.
Here, however, we employ the polaron transformation in a different context:
It is used to recast interactions in the extended system, between the spin and the reaction coordinates.  
We emphasize that we are not computing here the heat current in the polaron frame. Notably, only by inspecting the  polaron-transformed Hamiltonian we gain insight into the two transport contributions.

We apply a small-polaron transformation to the left reservoir in Eq. (\ref{eq:HSBRC}) $\hat{\tilde{H}}_{SB-RC} = \hat{U}_P\hat{H}_{SB-RC}\hat{U}_P^{\dagger}$ with the unitary transformation \cite{Mahan}
\bea
    \hat{U}_P^L = e^{\frac{\lambda_L}{\Omega_L}\hat{\sigma}_x (\hat{a}_L^{\dagger} - \hat{a}_L)}.
    \label{eq:UPL}
\eea
This results in 
%
\begin{widetext}
\bea
\nonumber
\hat{\Tilde{{H}}}_{SB-RC} &=& \frac{\Delta}{4}\left[ \left(\hat{\sigma}_z
-i\hat{\sigma}_y\right)
e^{\frac{\lambda_L}{\Omega_L}(\hat{a}_L^{\dagger} - \hat{a}_L)} +
\left(\hat{\sigma}z+ i\hat{\sigma}_y\right)
 e^{-\frac{\lambda_L}{\Omega_L}(\hat{a}_L^{\dagger} - \hat{a}_L)} \right] 
+\Omega_R \hat{a}_R^{\dagger}\hat{a}_R + \Omega_L \hat{a}_L^{\dagger}\hat{a}_L   \nonumber\\
&+&\lambda_R \sin\theta (\hat{a}_R^{\dagger} + \hat{a}_R)\hat{\sigma}_x 
+ \lambda_R \cos\theta (\hat{a}_R^{\dagger} + \hat{a}_R) \frac{1}{2}
\left[ \left(\hat{\sigma}_z
-i\hat{\sigma}_y\right)
e^{\frac{\lambda_L}{\Omega_L}(\hat{a}_L^{\dagger} - \hat{a}_L)} +
\left(\hat{\sigma}z+ i\hat{\sigma}_y\right)
 e^{-\frac{\lambda_L}{\Omega_L}(\hat{a}_L^{\dagger} - \hat{a}_L)} \right] 
\nonumber\\ 
&-& \frac{2\lambda_L}{\Omega_L} \hat{\sigma}_x \sum_k g_{k,L}(\hat{b}^{\dagger}_{k,L} + \hat{b}_{k,L}) 
+ (\hat{a}_L^{\dagger} + \hat{a}_L - \frac{2\lambda_L}{\Omega_L}\hat{\sigma}_x)^2 \sum_{k} \frac{g_{k,L}^2}{\omega_{k,L}} + (\hat{a}_R^{\dagger} + \hat{a}_R)^2 \sum_{k} \frac{g_{k,R}^2}{\omega_{k,R}} 
 \nonumber\\
&+& \sum_{\alpha} (\hat{a}_\alpha^{\dagger} + \hat{a}_\alpha) \sum_k g_{k,\alpha} (\hat{b}_{k,\alpha}^{\dagger} + \hat{b}_{k,\alpha})
+ \sum_{k,\alpha} \omega_{k,\alpha}  \hat{b}_{k,\alpha}^{\dagger} \hat{b}_{k,\alpha}.
\label{eq:HSBRCP}
\eea
\end{widetext}
%
The terms in Eq. (\ref{eq:HSBRCP}) are organized as follows: The first line includes the contribution from the spin and the two bare reaction coordinates. 
At low temperature when the RCs are in their ground state, the impact of the polaron dressing is to renormalize the spin splitting $\Delta\to$
 $\Delta e^{-\frac{\lambda^2}{2\Omega^2}}$ \cite{Nick2021}. 

The second line includes the interaction between the right RC and the spin, part of it is unaltered from the mapping since $[\hat{U}_P,\hat{\sigma}_x] = 0$,
while the second part  arises due to  $[\hat{U}_P,\hat{\sigma}_z] \neq 0$. These two terms are central to uncovering transport mechanisms, and will be discussed below.
The third line includes the  coupling between the spin and the left residual reservoir, as well as the quadratic term, which represents the reorganization energy of the reservoirs. 
The last line describes the coupling between the RCs and the residual baths, including the bath Hamiltonians themselves.
We now separately analyze two limiting models, $\theta = \pi/2$ and $\theta = 0$.

{\it (i) $\sigma_x-\sigma_x$ model.}
When $\theta = \pi/2$, the model Eq. (\ref{eq:HSB}) reduces to the 
$\sigma_x-\sigma_x$ model. In this symmetric case, it is beneficial to perform an additional polaron transformation on the right side, applying $\hat U_P^R$, in a complete analogy to Eq. (\ref{eq:UPL}).
We end up with ($\alpha=L,R$ and ignoring constant terms),
\bea
&&\hat{\Tilde{{H}}}_{SB-RC}^{\sigma_x-\sigma_x} =
\nonumber\\
&&\frac{\Delta}{4}\left[
 \left(\hat{\sigma}_z
-i\hat{\sigma}_y\right) e^{\sum_{\alpha}\frac{\lambda_{\alpha}}{\Omega_{\alpha}}(\hat{a}_{\alpha}^{\dagger} - \hat{a}_{\alpha})} +  \left(\hat{\sigma}_z
+i\hat{\sigma}_y\right) e^{-\sum_{\alpha}\frac{\lambda_{\alpha}}{\Omega_{\alpha}}(\hat{a}_{\alpha}^{\dagger} - \hat{a}_{\alpha})} \right]
\nonumber\\ 
&&+\Omega_R \hat{a}_R^{\dagger}\hat{a}_R + \Omega_L \hat{a}_L^{\dagger}\hat{a}_L
+\sum_{\alpha}(\hat{a}_{\alpha}^{\dagger} + \hat{a}_{\alpha} - \frac{2\lambda_{\alpha}}{\Omega_{\alpha}}\hat{\sigma}_x)^2 \sum_{k} \frac{g_{k,{\alpha}}^2}{\omega_{k,{\alpha}}} 
 \nonumber\\
&&- \sum_{\alpha} \frac{2\lambda_{\alpha}}{\Omega_{\alpha}}\hat{\sigma}_x \sum_{k} g_{k,\alpha} (\hat{b}_{k,\alpha}^{\dagger} + \hat{b}_{k,\alpha}) \nonumber \\
&&+\sum_{\alpha} (\hat{a}_\alpha^{\dagger} + \hat{a}_\alpha) \sum_k g_{k,\alpha} (\hat{b}_{k,\alpha}^{\dagger} + \hat{b}_{k,\alpha})
+ \sum_{k,\alpha} \omega_{k,\alpha}  \hat{b}_{k,\alpha}^{\dagger} \hat{b}_{k,\alpha}
\label{eq:HPxx}
\eea
To the lowest order in $\lambda$, we get (ignoring constant shifts),
\bea
&&\hat{\Tilde{{H}}}_{SB-RC}^{\sigma_x-\sigma_x} =
\frac{\Delta}{2}
\hat{\sigma}_z  - \frac{\Delta}{2}i \hat{\sigma}_y \sum_{\alpha} \frac{\lambda_{\alpha}}{\Omega_{\alpha}}(\hat{a}^{\dagger}_{\alpha} - \hat{a}_{\alpha})
+\Omega_R \hat{a}_R^{\dagger}\hat{a}_R + \Omega_L \hat{a}_L^{\dagger}\hat{a}_L
\nonumber\\
&&+
\sum_{\alpha}(\hat{a}_{\alpha}^{\dagger} + \hat{a}_{\alpha})  
\left[
\hat{\sigma}_x \left(\frac{-2\lambda_{\alpha}}{\Omega_{\alpha}}\right)
\sum_{k} \frac{g_{k,{\alpha}}^2}{\omega_{k,{\alpha}}} 
+\sum_k g_{k,\alpha} (\hat{b}_{k,\alpha}^{\dagger} + \hat{b}_{k,\alpha})
\right] 
 \nonumber\\
 &&- \sum_{\alpha} \frac{2\lambda_{\alpha}}{\Omega_{\alpha}}\hat{\sigma}_x \sum_{k} g_{k,\alpha} (\hat{b}_{k,\alpha}^{\dagger} + \hat{b}_{k,\alpha}) \nonumber\\
 &&+ 
  \sum_{k,\alpha} \omega_{k,\alpha}  \hat{b}_{k,\alpha}^{\dagger} \hat{b}_{k,\alpha}+\sum_{\alpha}(\hat{a}_{\alpha}^{\dagger} + \hat{a}_{\alpha})^2  \sum_{k} \frac{g_{k,{\alpha}}^2}{\omega_{k,{\alpha}}}.
\label{eq:HPxx}
\eea
Assuming for simplicity that temperature is not high enough to excite the RCs, it is clear from the third line that heat transport takes place through the system's current:
The left hot bath excites the the spin with the 
transition $\hat \sigma_x$, and this system's operator is further coupled to the right residual bath. Eq. (\ref{eq:HPxx}) in fact provides the foundation for the effective RC model described in Ref. \cite{Nick2021}.
At weak coupling, the $\sigma_x-\sigma_x$ model thus describes sequential heat transport, from bath to spin, and spin to bath, 
and it only accounts for system's current.

{\it (ii) $\sigma_x-\sigma_z$ model.}
We return to Eq. (\ref{eq:HSBRCP}), but now use $\theta = 0$. 
After expanding the exponents to lowest order in $\lambda$, the Hamiltonian reduces to 
\bea
\nonumber
&&\hat{\Tilde{{H}}}_{SB-RC}^{\sigma_x-\sigma_z} = \frac{\Delta}{2} \hat{\sigma}_z - \frac{i \lambda_L \Delta}{2\Omega_L}\hat{\sigma}_y(\hat{a}^{\dagger}_L - \hat{a}_L) + \sum_{k,\alpha} \omega_{k,\alpha}  \hat{b}_{k,\alpha}^{\dagger} \hat{b}_{k,\alpha}
\nonumber \\
&&+\Omega_R \hat{a}_R^{\dagger}\hat{a}_R + \Omega_L \hat{a}_L^{\dagger}\hat{a}_L + \lambda_R \hat{\sigma}_z  (\hat{a}^{\dagger}_R + \hat{a}_R) 
\nonumber\\
&&-i\frac{ \lambda_R\lambda_L}{\Omega_L}
  (\hat{a}_R^{\dagger} + \hat{a}_R) 
\hat{\sigma}_y (\hat{a}_L^{\dagger} - \hat{a}_L) 
\nonumber\\ 
&&- \frac{2\lambda_L}{\Omega_L} \hat{\sigma}_x \sum_k g_{k,L}(\hat{b}^{\dagger}_{k,L} + \hat{b}_{k,L}) 
+ (\hat{a}_L^{\dagger} + \hat{a}_L - \frac{2\lambda_L}{\Omega_L}\hat{\sigma}_x)^2 \sum_{k} \frac{g_{k,L}^2}{\omega_{k,L}} 
 \nonumber\\
&&+ (\hat{a}_R^{\dagger} + \hat{a}_R)^2 \sum_{k} \frac{g_{k,R}^2}{\omega_{k,R}} 
+\sum_{\alpha} (\hat{a}_\alpha^{\dagger} + \hat{a}_\alpha) \sum_k g_{k,\alpha} (\hat{b}_{k,\alpha}^{\dagger} + \hat{b}_{k,\alpha}).
\label{eq:Hxzweak}
\eea
%
To the lowest order in the interaction energy, we identify the term responsible for heat transport as
 $\frac{-i \lambda_R\lambda_L}{\Omega_L} (\hat{a}_R^{\dagger} + \hat{a}_R)\hat{\sigma}_y(\hat{a}_L^{\dagger} - \hat{a}_L) $. It captures the main physics:
Heat is exchanged in the $\sigma_x-\sigma_z$ model
due to the two reaction coordinates becoming effectively coupled through the spin,
an effect reminiscent of ``superexchange" for charge current \cite{Scholes03,Segal11}. 
Since the RCs are in fact part of the heat baths, this term 
corresponds to the direct, inter-bath current as described by Eq. (\ref{eq:current}). 
%

To organize our numerical results and polaron analysis: 
There are two types of transport pathways. The system's current arises due to internal transitions in the spin. 
At weak coupling, this contribution can be captured by standard second-order QME approaches. 
On the other hand, the inter-bath current arises due to transitions in the RCs (bath), with inter-bath transitions mediated by the spin.
This form of current is immediately of higher order in the system-bath coupling strength as its prefactor is $\lambda_{L}\lambda_R$. To simulate it one needs to use methods capturing effects beyond second order transport characteristics. 
In what follows, we present simulations of the steady state heat current using the RC-QME method.

\begin{figure*}[ht]
 \center
 \includegraphics[width=2\columnwidth]{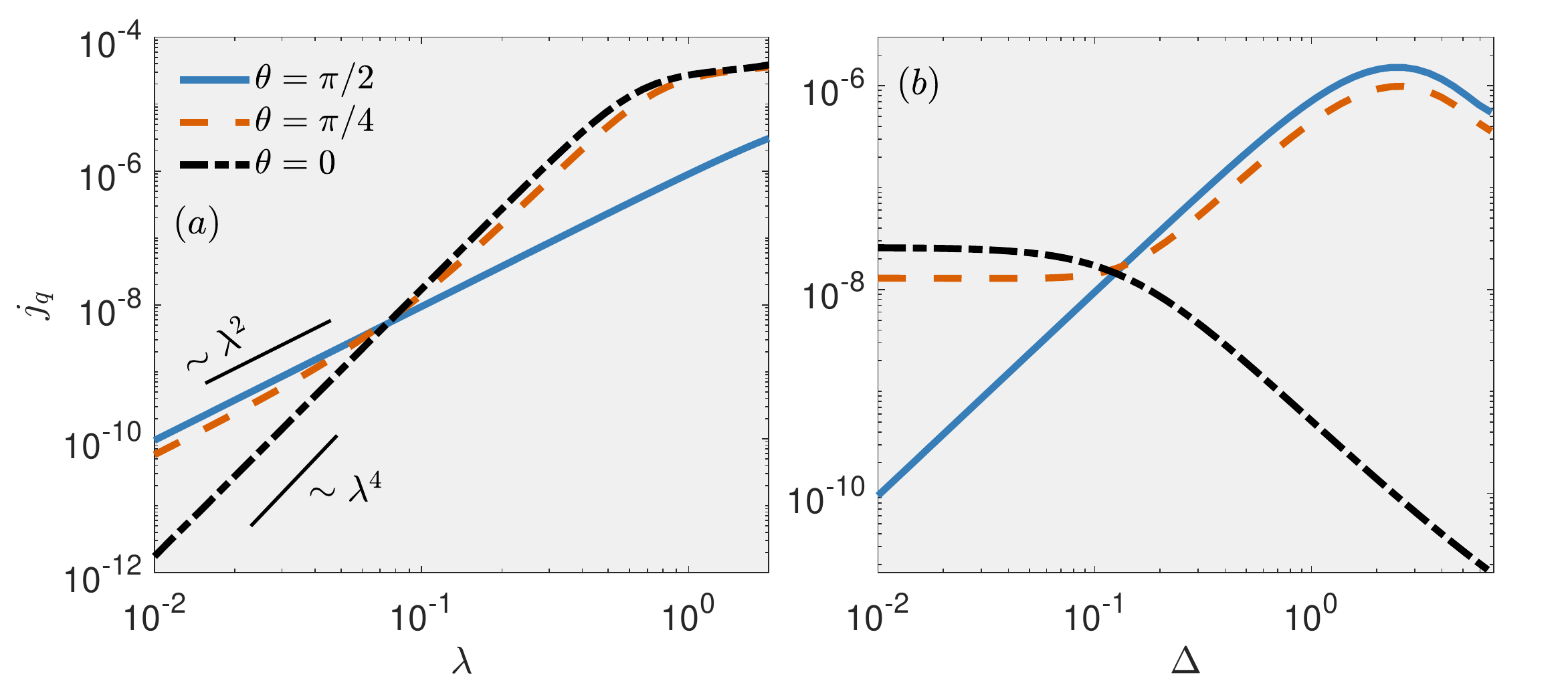} 
\caption{Steady state heat current as a function of (a) coupling strength $\lambda$ and (b) spin splitting $\Delta$. 
The currents are plotted for three different angles, $\theta = \pi/2$ (solid), $\theta = \pi/4$ (dashed), and $\theta = 0$ (dashed-dotted). 
The dark solid lines in panel (a) indicate the scaling behavior of the system's current ($\lambda^2$) and the inter-bath current ($\lambda^4$). 
Parameters are $\Omega = 10$, $\gamma = 0.0071/\pi$, $T_h = 1$, $T_c = 0.5$, $\Lambda = 1000$, $M = 4$ as well as $\Delta = 0.1$, (a) and $\lambda = 0.1$ (b).}
\label{fig:Jq_lam}
\end{figure*}

\begin{figure*}[hbt]
 \center
 \includegraphics[width=2\columnwidth]{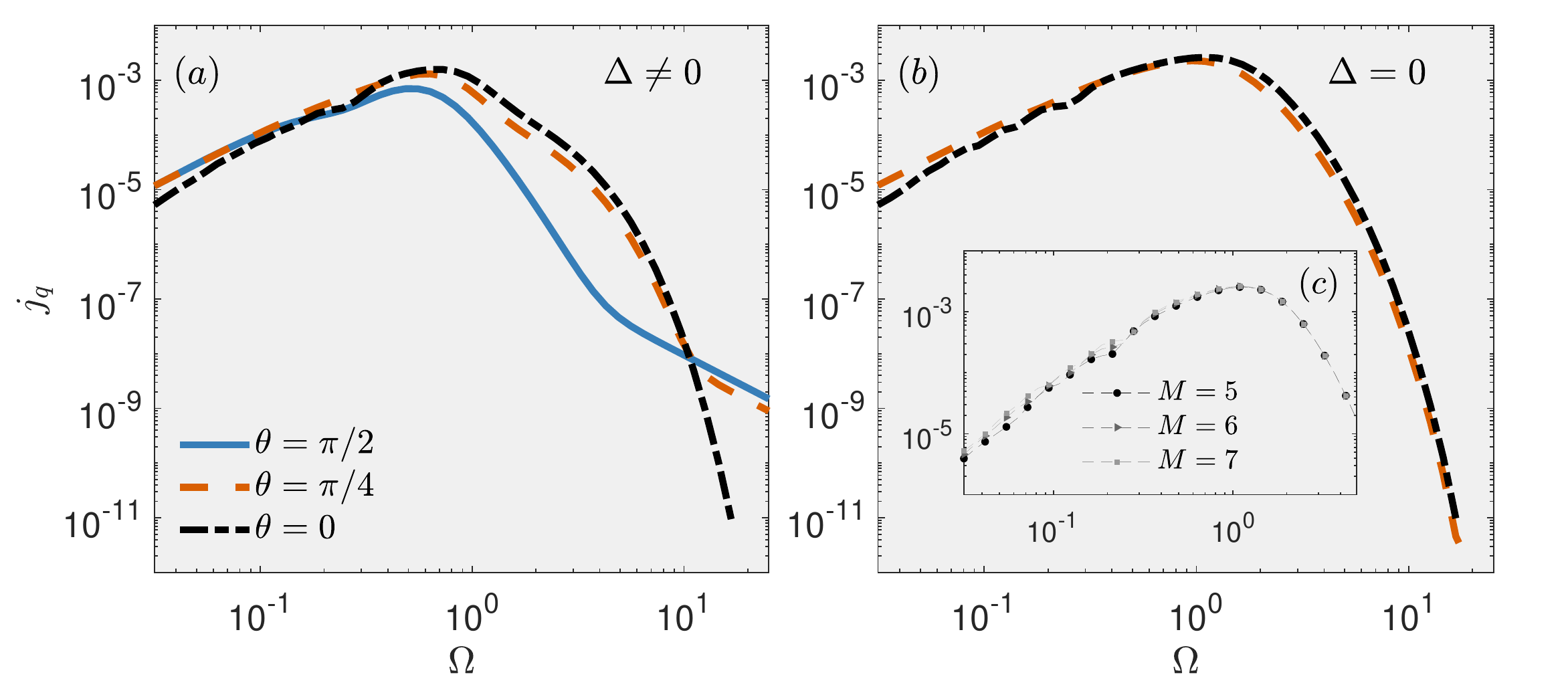} 
\caption{Steady state heat current as a function of the RC frequency, $\Omega$, 
plotted with three orientation angles
$\theta = \pi/2$ (solid), $\theta = \pi/4$ (dashed), and $\theta = 0$ (dashed-dotted).
(a) $\Delta =0.1$ and $\lambda=0.1$. 
(b) $\Delta = 0$ and $\lambda=0.1$. 
Other parameters are the same as in Figure \ref{fig:Jq_lam}, 
except $M=7$. 
Panel (c) shows the convergence of the steady state current 
with $M = 5$ (circle), $M=6$ (triangle) and $M=7$ (square) for the case of $\Delta = 0$.}
\label{fig:Jq_Om}
\end{figure*}

\subsection{RC-Quantum Master Equation Method}
We outline the principles of the RC-QME method, which  we employ next to
study heat current in the RC-SB Hamiltonian, Eq. (\ref{eq:HSBRC}).
For more details, we refer readers to Ref. \cite{Nick2021}.
The key point is that after the reaction coordinate mapping is performed, 
one assumes that the coupling between the RCs and the residual baths is weak. 
This amounts to $\gamma_{\alpha} \ll 1$;  the Brownian
spectral function is assumed narrow.
As a result, one can employ the standard Born-Markov QME method 
on the {\it extended system} and study effects beyond second order 
in $\lambda$, yet in a computationally cheap manner.

Employing the RC-QME method follows three steps: 
(i) We truncate the reaction coordinates and include $M$ levels each. 
The  value of $M$ is chosen large enough to converge numerical simulations with respect to this parameter. 
After the truncation, the extended system is of an $2\times M^2$ dimension.
(ii) The  truncated SB-RC Hamiltonian is diagonalized numerically. 
The operators of the system that are coupled to the baths are transformed into the diagonal representation.
(iii) The Redfield equation is solved in steady state in the energy eigenbasis of the extended system of the SB-RC Hamiltonian. 
The dynamics is assumed Markovian, and the initial state of the open system is a product state of the extended system and baths, the latter prepared in canonical states at a given temperature. 
We further ignore the principal values (imaginary terms) in the Fourier-transformed 
bath-bath correlation functions, see Appendix A. 

Formally, the Redfield equation is given by $\dot \rho_{ES}(t)=-i[\hat H_{ES}^{D},\rho_{ES}] + \sum_{\alpha}{\cal D}_{\alpha}(\rho_{ES})$
\cite{Nitzan}. The dissipators to the residual baths, 
${\cal D}_{\alpha}(\rho_{ES})$ are additive, 
given the weak coupling approximation of the RCs to their residual baths. However, the dissipators depend in a {\it nonadditive} manner on the original coupling parameters, $\lambda_{\alpha}$. 
We solve the Redfield equation in the energy basis under the constraint of $\Tr[\rho_{ES}(t)] = 1$ and obtain the steady state density matrix of the extended system, $\rho_{ES}^{ss}$. The steady state heat current at the $\alpha$ contact is given by $j_{\alpha}={\rm Tr}\left[ {\cal D}_{\alpha}(\rho_{ES}^{ss}) \hat H_{ES}^{D}\right]$; the heat current is defined positive when flowing from the bath towards the system.

\subsection{Simulations}
\label{Sec:Simul}

We use the RC-QME method and simulate the steady state heat current.
Our objective is to exemplify and pinpoint the differences between the system and inter-bath transport mechanisms.
For simplicity, we consider symmetric RC parameters: $\lambda = \lambda_L = \lambda_R$, $\Omega = \Omega_L = \Omega_R$, and $\gamma = \gamma_L = \gamma_R$. 

Figure \ref{fig:Jq_lam} (a) manifests the scaling properties of the steady state heat current with the coupling strength $\lambda$, in the weak coupling regime. 
We use three different angles $\theta$. 
The solid line ($\theta = \pi/2$) corresponds to the case where the two system operators commute ($\sigma_x-\sigma_x$ model).
According to  Eq. (\ref{eq:current}) and the analysis of Sec. \ref{Sec:mech}, in this situation at weak coupling  we should observe the system's current, which is mediated by transitions in the spin, as the sole mechanism of transport. This current scales as $j_q \propto \lambda^2$.

On the contrary, the solid-dashed line ($\theta=0$) corresponds to the other extreme, where the system Hamiltonian and the right system operator commute ($\sigma_x-\sigma_z$ model).
As a result, the system's current nullifies and heat 
transport must go beyond second order in the system-bath coupling parameter.
This is shown by the scaling relation $j_q \propto \lambda^4$, which corresponds to the inter-bath current as described in Sec. \ref{Sec:mech}.

The dashed line ($\theta = \pi/4$) pictures an 
intermediate case, when both system's and inter-bath currents contribute at weak coupling. 
As a result, at the ultra-weak coupling the current scales as 
$\lambda^2$ (system's current), but as the coupling is enhanced (yet it is still maintained small) there is a
smooth transition, and the inter-bath current takes over showing the scaling $j_q\propto \lambda^4$.
These scaling relations, and the turnover behavior
for the intermediate angle $\theta=\pi/4$ were demonstrated in Ref. \cite{Junjie20} by using the extended hierarchical equations of motion (HEOM) technique \cite{Tanimura20} (albeit using the super-Ohmic spectral function for the baths). It is significant to confirm here that the RC-QME method, a rather economical and transparent technique, captures the correct scaling behavior and the continuous transition between them.

To strengthen the argument that the system's current arises from transitions caused in the spin, 
while the inter-bath current is a reservoir effect, we show in Figure \ref{fig:Jq_lam} (b) the steady state current as a function of the spin splitting $\Delta$. We find that for $\theta = \pi/2$ the current diminishes as the spin splitting goes to zero, confirming the critical role of transitions in the spin system (in sequential transport, which is the dominant system's current at weak coupling, heat is transported in quanta of $\Delta$).
On the contrary, when $\theta = 0$ we observe a saturation of the current as the splitting approaches zero. This supports the point that in this model heat current  is caused by inter-bath effects. Although, we note that the spin still plays a role in this transport pathway as there is still non-trivial dependence on the splitting as $\Delta$ varies. 
Finally, similarly to Figure \ref{fig:Jq_lam} (a), for the intermediate  model ($\theta = \pi/4$) the current  smoothly transitions between the two mechanisms, based on which pathway provides a larger magnitude in heat current.
The $\Delta$ dependence of the energy current was studied in Ref. \cite{Junjie20} using the extended HEOM method. It is substantial to note that the RC-QME method can capture the correct behavior.

Next, to further confirm the inter-bath nature of transport in the $\sigma_x-\sigma_z$ model,
we investigate the dependence of the heat current 
on $\Delta$ and the RC frequency $\Omega$. Figure \ref{fig:Jq_Om} shows the steady state heat current as a
 function $\Omega$ for the same three angles used in Figure \ref{fig:Jq_lam}. 
In Figure \ref{fig:Jq_Om} (a) we study behavior for $\Delta \neq 0$, 
while in Figure \ref{fig:Jq_Om} (b) we use $\Delta=0$. 
Our main observations are:
(i) The spin-boson model ($\theta = \pi/2$) supports current only when $\Delta \neq 0$, as is expected from system's current (this current is null in panel (b)).
(ii) The current flowing  in the $\sigma_x-\sigma_z$ model ($\theta = 0$) 
barely changes between the two panels. This is expected as heat flows in this model arises from transitions of the RCs (reservoirs). 
When increasing $\Omega$,
the inter-bath current grows, reaches a maximum, and decays to zero for large $\Omega$. This is because as $\Omega$ grows, the reaction coordinate levels become inaccessible to be thermally populated, resulting in a suppressed heat current. 
We therefore find that 
the role of $\Omega$ in the inter-bath transport is analogous to the role of $\Delta$ for the system's current. Namely, once $\Omega$ becomes approximately resonant with the temperature of the baths, inter-bath transport is maximized. When $\Omega$ is large and excited states cannot be thermally excited, a rapid suppression of the current is observed. 
Figure \ref{fig:Jq_Om} panel (c) exemplifies the convergence of the 
heat current with respect to the number of RC levels, $M$. 
We notice that once $\lambda / \Omega < 1$, convergence is achieved.
Surprisingly, however, even in situations where $\lambda / \Omega \approx 1$, 
convergence with respect to $M$ is still good: 
The overall trend of an increase in the magnitude of the current with $\Omega$ is followed.

Before  moving on to the three-level ladder model, we discuss in more details the relation of our work to Ref. \cite{Junjie20}. There, the heat current characteristics in the generalized spin-boson model were studied using the numerically exact extended HEOM technique. 
Using that method, different scaling relations of the heat current were observed, as we also show in Figure \ref{fig:Jq_lam}.
Furthermore, the nonequilibrium polaron-transformed Redfield equation (NE-PTRE) \cite{CaoNEPTRE} was implemented in Ref. \cite{Junjie20} to explain the behavior of the current in the $\sigma_{x}-\sigma_{z}$ model at $\Delta\to0$. It should be stressed that the polaron transformation was performed there on the original, pre-RC mapped model encompassing all bath modes, Eq. (\ref{eq:HSB}). In contrast, here we perform the polaron transformation {\it after} the RC mapping, and only on the individual RCs.
Ref. \cite{Junjie20} further focused on the enhancement that the non-commuting system operators provide to the steady state heat flux and to the thermal rectification effect, which we do not examine here.

In our work here, in contrast, we focus on physical mechanisms underling the two contributions to the current. We aim to establish the RC-QME method, and we show that the RC-mapped Hamiltonian, in particular after the polaron transformation over the RCs, offers a transparent starting point for studying transport mechanisms. 

Compared to the extended HEOM, which is numerically-exact, the RC-QME offers cheap computations and a deeper understanding.
Compared to the NE-PTRE method, the RC-QME offers more generality: It is cumbersome to perform the polaron transformation on multiple baths for non-commuting operators, the starting point of the NE-PTRE method. Thus, unless the model is symmetric, the NE-PTRE can handle strong coupling at one bath only.
In contrast, extracting degrees of freedom from several thermal baths and adding them to the system can be readily done for general coupling models with different $\theta$, and for multiple baths making the RC-QME method a useful tool in quantum thermodynamics \cite{Nick2021,QAR-Felix}. 



\section{Heat transport in the ladder model}
\label{Sec:ladder}

\subsection{Model}
Building on our knowledge of different transport mechanisms,
we now propose a three-level, ladder quantum system. Here, at weak coupling, heat current takes place solely due to inter-bath processes, rather than through the excitation of the system.
The goal of this section is to bring forward this unique model. We show that
the current scales as $j_q \propto \lambda^4$ at weak system-bath coupling 
$\lambda$, and that it can be largely tuned by modifying the spectral properties of the bath (frequency $\Omega$).
The ladder model is given by the following Hamiltonian ($\alpha=L,R$),
\bea
    \hat{H}_{LD} = &&\sum_{i=0}^{2} \epsilon_i \ket{i}\bra{i} + \sum_{k,\alpha} \hat{S}_{\alpha}^2 \frac{f_{k,\alpha}^2}{\nu_{k,\alpha}} + 
    \sum_{k,\alpha} \hat{S}_{\alpha} f_{k,\alpha} (\hat{c}_{k,\alpha}^{\dagger} + \hat{c}_{k,\alpha}) 
    \nonumber \\ 
    &+& \sum_{k,\alpha} \nu_{k,\alpha} \hat{c}_{k,\alpha}^{\dagger} \hat{c}_{k,\alpha}.
    \label{eq:H3}
\eea
Here, $\epsilon_i$ is the energy of the $i$th level of the ladder. 
For simplicity, we set the ground state at $\epsilon_0 = 0$, define $\epsilon_1 = \Delta$ and fix $\epsilon_2 = 1$.  $f_{k,\alpha}$ denotes the coupling strength of an oscillator mode in the $\alpha$th bath with frequency $\nu_{k,\alpha}$ to the system. 
The two baths couples to two distinct transitions via the system operators $\hat S_H = \ket{1}\bra{2} + h.c.$ and $\hat S_C = \ket{0}\bra{1} + h.c.$. 
The setup is depicted in Figure \ref{fig:Diagram} (b).
The system-bath interactions are captured by a Brownian spectral density functions, as indicated in Eq. (\ref{eq:JBrownian}). 

At weak coupling at the level of the 
second-order BMR QME, the ladder model cannot conduct heat in steady state. 
We derive this result in the Appendix. 
However, in this limit the ladder model conducts through inter-bath transitions, 
and it therefore scales as $j_q\propto \lambda^4$.
To study this effect, 
We once again must implement the RC transformation---in analogy to what was done for the generalized spin-boson model.
After extracting one oscillator from each reservoir we reach the following Hamiltonian,
\bea
    \nonumber
    \hat{H}_{LD-RC} &=& \sum_{i=0}^{2} \epsilon_i \ket{i}\bra{i} + \sum_{\alpha} \hat{S}_\alpha^2 \frac{\lambda_{\alpha}^2}{\Omega_{\alpha}} + \sum_{\alpha} \Omega \hat{a}_{\alpha}^{\dagger} \hat{a}_{\alpha}
    \\ \nonumber
    &+& \sum_{\alpha} \lambda_\alpha \hat{S}_{\alpha} (\hat{a}_{\alpha}^{\dagger} + \hat{a}_{\alpha}) + \sum_{k,{\alpha}} \frac{g^2_{k,{\alpha}}}{\omega_{k,{\alpha}}} (\hat{a}_{\alpha}^{\dagger} + \hat{a}_{\alpha})^2  
    \\ \nonumber
    &+& \sum_{\alpha} (\hat{a}_{\alpha}^{\dagger} + \hat{a}_{\alpha}) \sum_k g_{k,{\alpha}} (\hat{b}_{k,{\alpha}}^{\dagger} + \hat{b}_{k,{\alpha}}) + \sum_{k,{\alpha}}  \omega_{k,{\alpha}} \hat{b}_{k,{\alpha}}^{\dagger} \hat{b}_{k,{\alpha}}.
    \\
    \label{eq:H3RC}
\eea
The coupling between the RCs and the residual baths are given by Ohmic spectral density functions post mapping, see Eq. (\ref{eq:JOhmic}). 
Furthermore, all parameters carry the same meaning as in the generalized spin-boson Hamiltonian, Eq. (\ref{eq:HSBRC}). In particular, $\lambda$ denotes the system-bath coupling strength as it appears in the spectral function Eq.  (\ref{eq:JBrownian}).
\subsection{Simulations}

\begin{figure}
\centering
\includegraphics[width=1\columnwidth]{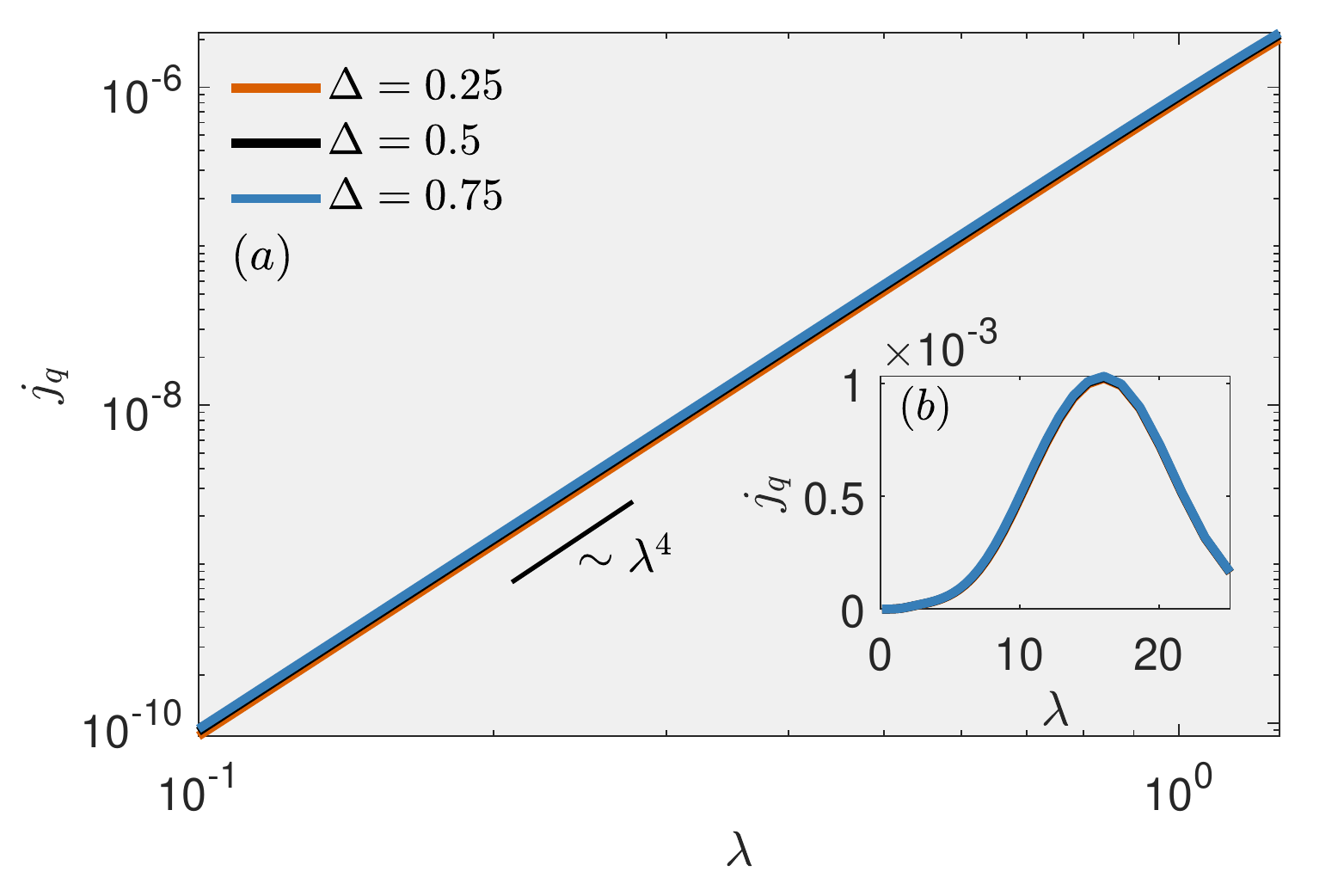} 
\caption{Steady-state heat current in the three-level ladder model as a function of coupling parameter $\lambda$ on (a) a logarithmic scale and (b) a linear scale for three different splittings, $\Delta = 0.25$ (orange), $\Delta = 0.5$ (black), and $\Delta = 0.75$ (blue). 
The dark solid line in panel (a) indicates the scaling behavior of the inter-bath current. 
Parameters are $\epsilon_0 = 0$, $\epsilon_2 = 1$, 
$\Omega = 10$, $\gamma = 0.0071/\pi$, $T_h = 1$, $T_c = 0.5$, $\Lambda = 1000$, $M=5$.
Panel (b) illustrates the suppression of the heat current at strong coupling. 
    }
    \label{fig:3lvl_lambda}
\end{figure}

We simulate the heat current using the Redfield QME, after a reaction coordinate transformation is applied, Eq. (\ref{eq:H3RC}). 
We take the RC parameters equal in numerical simulations, $\lambda = \lambda_{\alpha}$, and $\Omega = \Omega_{\alpha}$. 
Figure \ref{fig:3lvl_lambda} displays the heat current as a function of the coupling strength, $\lambda$, for three different values of $\Delta$ (the position of the intermediate level).
Results are presented on a logarithmic scale, exposing the scaling relation in the weak coupling regime, $j_q \propto \lambda^4$, characteristic to the inter-bath current. 
The absence of a lower order scaling confirms that the inter-bath current 
is the leading mechanism of transport at weak coupling.
We also find that the value of $\Delta$ has 
an insignificant impact on the current, supporting the argument that heat exchange between baths 
is the dominant transport pathway.
%
%

So far, we focused on the weak coupling limit with $\lambda\ll \Omega$. 
Complementing this, we further demonstrate in Figure \ref{fig:3lvl_lambda} (b) 
the behavior of he heat current as we push $\lambda$ into the strong coupling regime,
$\lambda\gtrsim\Omega$ \cite{comment3}. 
The heat current rises with $\lambda$, reaches a maximum, and then decays. 
This trend of the ladder model is parallel to the behavior of the spin-boson model \cite{Nick2021}, 
albeit here originating from the inter-bath current.
Results beyond $\lambda=10$  were not fully converged, and they required simulations with higher $M$ values,
which posed a computational challenge. 
These results are presented to provide the broad picture of transport in the model.
Nevertheless, while the magnitude of the current was still changing with $M$, trends and
the peak position stayed intact.


\begin{figure}
    \centering
  \includegraphics[width=1\columnwidth]{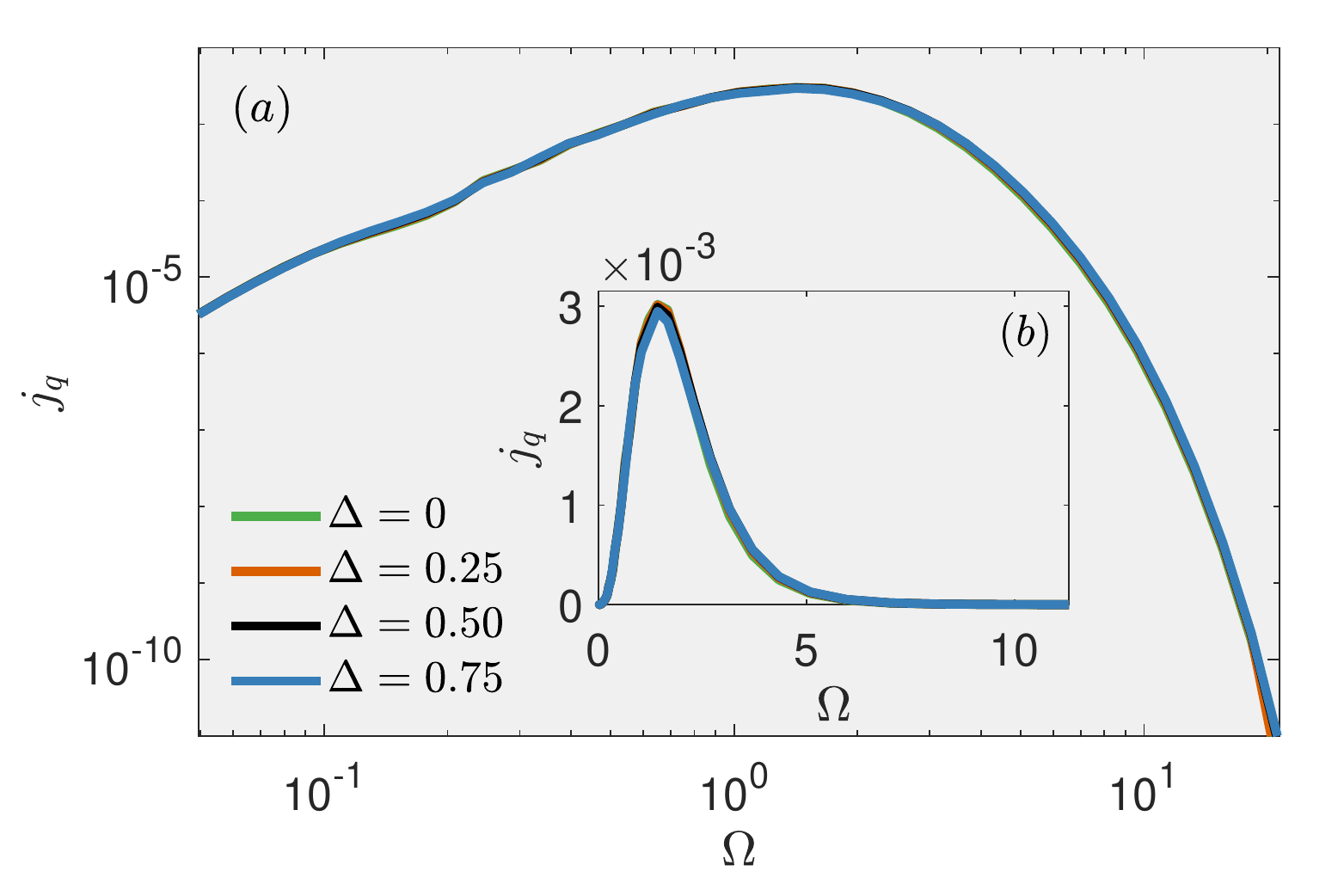}
\caption{Steady-state heat current in the three-level ladder model as a function of the RC frequency $\Omega$ for $\Delta = 0$ (green), $\Delta = 0.25$ (orange), $\Delta = 0.5$ (black), and $\Delta = 0.75$ (blue), showing that the current does not depend on the internal energetics, $\Delta$.
a) The current on logarithmic and (b) linear scales.
Other parameters are identical to Figure \ref{fig:3lvl_lambda}, with $\lambda = 1$ and 
$M=5$. }
    \label{fig:3lvl_Omega}
\end{figure}

To further illustrate that heat flow in the ladder model 
is due to the inter-bath mechanism, we simulate the heat current as a function 
of the reaction coordinate frequency, $\Omega$ in Figure \ref{fig:3lvl_Omega}. 
We observe similar trends as found in Figure \ref{fig:Jq_Om}. 
Namely, an increase of the current at small $\Omega$ followed by a rapid suppression as $\Omega$ grows. 
This suppression is attributed to the small population of excited states of the RC 
as one increases their frequency,  thereby reducing the inter-bath current. 
Testing convergence of results is performed analogously
to Figure \ref{fig:Jq_Om}. 
Beyond  $\Omega \approx \lambda$, results are converged whereas for $\Omega < \lambda$ 
the magnitude of the current was still visibly changing with $M$ on the presented scale.

\section{Summary}
\label{Sec:Conclusion}

The goal of this work has been threefold:

(i) To establish the RC mapping and the RC-QME method as an effective 
tool for studying steady-state quantum transport characteristics.
This study goes beyond earlier
benchmarking that focused on the equilibration problem with respect to a single heat bath \cite{Anders},
as well as previous studies that demonstrated energy renormalization 
of the {\it system's current} as a key signature of strong system-bath couplings \cite{Nick2021,QAR-Felix}.
In contrast, here we demonstrate that more fundamentally, 
the exact RC mapping (as well as the approximate RC-QME implementation) 
can realize two distinct transport mechanisms: a system's current, 
which at weak coupling reduces to sequential transport, and an inter-bath current, 
with bath modes directly exchanging energy between them, and the quantum system acting as a bridge.

(ii) To realize and tune different heat  transport mechanisms.
Significantly, we gained understanding on transport mechanisms and their characteristics without explicitly performing transport calculations, only by inspecting the RC-mapped Hamiltonian, especially after doing an additional polaron transformation.

(iii) To bring forward models that  exclusively enact a direct-bath current at weak coupling. 
Besides the $\sigma_x-\sigma_z$ model, which was introduced before,  we constructed and simulated the ladder model. We showed that heat current characteristics in the ladder model
are of an inter-bath nature at weak coupling with the current scaling quartically
with the system-bath coupling strength.

The main practical message of our work is that the reaction coordinate mapping technique captures nontrivial aspects of system-bath coupling: Beyond bath-induced levels renormalization  \cite{Nick2021,QAR-Felix}, the method captures inter-bath heat transfer effects.
As for physical results,
by inspecting the RC-mapped Hamiltonian numerically and analytically, and by performing
simulations extracting different scaling relations,
we identified two heat transport mechanisms: system's current and inter-bath current. 
The system's current is due to the baths thermally exciting and de-exciting the system, 
 with sequential transport dominating at weak coupling.
The inter-bath current corresponds to direct energy exchange between the modes of the baths, 
utilizing the system to build this exchange. 
These results were demonstrated in both the generalized spin-boson model and the proposed ladder model. 
The inter-bath current exists at weak coupling once the system is coupled to the bath with non-commuting operators, a prevalent scenario in  quantum thermodynamics. For example, in the commonly explored three-level quantum absorption refrigerator \cite{Luis} system's coupling operators do not commute, giving rise to inter-bath leakage currents \cite{QAR-Felix}. 

Combining the reaction coordinate mapping with a polaron transformation 
proved to prepare the Hamiltonian in a highly interpretive form, exposing transport pathways. 
Future work will be focused on studies of impurity dynamics at strong coupling using the RC-polaron mapping approach.


\begin{acknowledgments}
We acknowledge fruitful discussions with Junjie Liu.
DS acknowledges the NSERC discovery grant and the Canada Research Chair Program. 
The work of FI was partially funded by the Centre for Quantum Information and Quantum Control at the University of Toronto.
\end{acknowledgments}



\renewcommand{\theequation}{A\arabic{equation}}
\setcounter{equation}{0}  
\setcounter{section}{0} 
\section*{Appendix: Absence of system's current in the ladder model under the BMR QME}\label{app:1}


In this appendix, we focus on the ladder model at weak coupling.
We show that heat cannot be conducted in the system if the study is carried out using 
the second order Born-Markov Redfield QME, a method that only captures system's current.
Indeed, as we show in the main text, the heat current scales as $j_q\propto \lambda^4$ at weak coupling, pointing to the inter-bath current as the transport mechanism at small $\lambda$.

The ladder model is given by Eq. (\ref{eq:H3}).
In the Schr\"odinger picture, the BMR QME for the reduced density operator 
of the system is given by \cite{Nitzan},
\bea
\dot{\rho}_{mn}(t) &=& -i\omega_{mn} \rho_{mn}(t)
\nonumber\\ &-&
\sum_{\alpha}\sum_{j,k}[
R_{mj,jk}^{{\alpha}}(\omega_{kj}) \rho_{kn}(t) + R_{nk,kj}^{{\alpha}*}(\omega_{jk}) \rho_{mj}(t)
\nonumber\\ &-&
R_{kn,mj}^{{\alpha}}(\omega_{jm}) \rho_{jk}(t) - R_{jm,nk}^{{\alpha}*}(\omega_{kn}) \rho_{jk}(t)].
\label{eq: Redfield}
\eea
Here, $\omega_{mn}=\epsilon_m-\epsilon_n$ are the eigenenergies of the three-level system.
The elements of the $R$ super-operator are given by half Fourier transform
of bath autocorrelation functions,
\bea
R_{mn,jk}^{{\alpha}}(\omega) &=& \hat{S}^{\alpha}_{m,n} \hat{S}^{\alpha}_{j,k} \int_0^{\infty} d\tau e^{i\omega\tau} \langle \hat{B}_{\alpha}(\tau)\hat{B}_{\alpha}\rangle
\nonumber\\ &=&
\hat{S}^{\alpha}_{m,n} \hat{S}^{\alpha}_{j,k} [\Gamma_{{\alpha}}(\omega) + i\Delta_{{\alpha}}(\omega)].
\label{eq:dissipator}
\eea
These functions are evaluated with respect to the thermal state of their respective baths.
In what follows, we neglect the imaginary component of the dissipators, $\Delta_{\alpha}$, for simplicity. Working in the energy basis of the Hamiltonian Eq. (\ref{eq:H3}),
 the equations of motion for the populations of the reduced density matrix $\rho_{ii}(t) = \bra{i} \rho(t) \ket{i}$ become
\bea
\dot{\rho}_{00}(t) &=& -\left[R^C_{01,10}(\omega_{01}) + R^{*C}_{01,10}(\omega_{01})\right]\rho_{00}(t)
\nonumber\\
&+&\left[R^C_{10,01}(\omega_{10}) + R^{*C}_{10,01}(\omega_{10})\right]\rho_{11}(t)
\nonumber\\
\dot{\rho}_{11}(t) &=& -\left[R^C_{10,01}(\omega_{10}) + R^{*C}_{10,01}(\omega_{10})\right] \rho_{11}(t)
 \nonumber\\
&-& \left[R^H_{12,21}(\omega_{12}) + R^{*H}_{12,21}(\omega_{12})\right] \rho_{11}(t) 
 \nonumber\\
&+& \left[R^C_{01,10}(\omega_{01}) + R^{*C}_{01,10}(\omega_{01})\right]\rho_{00}(t) 
 \nonumber \\
&+& \left[R^H_{21,12}(\omega_{21}) + R^{*H}_{21,12}(\omega_{21})\right]\rho_{22}(t)
 \nonumber\\
\dot{\rho}_{22}(t) &=& -\left[R^H_{21,12}(\omega_{21}) + R^{*H}_{21,12}(\omega_{21})\right]\rho_{22}(t)
\nonumber \\
&+&\left[R^H_{12,21}(\omega_{12}) + R^{*H}_{12,21}(\omega_{12})\right]\rho_{11}(t)
\eea
These equation can be simplified.
Each pair of terms combine, and the half Fourier transform becomes a full integral. We identify those terms
as transition rates, providing the equations,
\bea
&\dot{\rho}_{00}(t)& = -k_{0 \xrightarrow[]{} 1}^C\rho_{00}(t) + k_{1 \xrightarrow[]{} 0}^C\rho_{11}(t)
 \nonumber\\
&\dot{\rho}_{11}(t)& = -(k_{1 \xrightarrow[]{} 0}^C + k_{1 \xrightarrow[]{} 2}^H) \rho_{11}(t) + k_{0 \xrightarrow[]{} 1}^C\rho_{00}(t) + k_{2 \xrightarrow[]{} 1}^H\rho_{22}(t)
 \nonumber\\
&\dot{\rho}_{22}(t)& = -k_{2 \xrightarrow[]{} 1}^H\rho_{22}(t) + k_{1 \xrightarrow[]{} 2}^H\rho_{11}(t)
\eea
Due to the natural decoupling between populations and coherences in this system, the above equation can be recast as a rate equation for the population of each state. In a compact notation, the kinetic-type equation is of the form $\dot{\Vec{p}} = \sum_{\alpha \in \{H,C\}} D_{\alpha} \Vec{p}$ with population
$\Vec{p} =(p_0 \, p_1 \,p_2)$ as diagonal elements of the system reduced density matrix and 
the dissipator matrices
\bea
D_C &=& 
\begin{pmatrix}
-k_{0 \xrightarrow[]{} 1}^C & k_{1 \xrightarrow[]{} 0}^C & 0\\
k_{0 \xrightarrow[]{} 1}^C & -k_{1 \xrightarrow[]{} 0}^C & 0\\
0 & 0 & 0
\end{pmatrix}
\\
D_H &=& 
\begin{pmatrix}
0 & 0 & 0\\
0 & -k_{1 \xrightarrow[]{} 2}^H & k_{2 \xrightarrow[]{} 1}^H\\
0 & k_{1 \xrightarrow[]{} 2}^H & -k_{2 \xrightarrow[]{} 1}^H
\end{pmatrix}
\eea
The steady-state populations are
\bea
{\Vec{p}}_{SS} = \frac{1}{\psi} \begin{pmatrix}
k_{1 \xrightarrow[]{} 0}^C k_{2 \xrightarrow[]{} 1}^H
\\
k_{0 \xrightarrow[]{} 1}^C k_{2 \xrightarrow[]{} 1}^H
\\
k_{0 \xrightarrow[]{} 1}^C k_{1 \xrightarrow[]{} 2}^H
\end{pmatrix},
\eea
where $\psi = k_{0 \xrightarrow[]{} 1}^C k_{1 \xrightarrow[]{} 2}^H + k_{0 \xrightarrow[]{} 1}^C k_{2 \xrightarrow[]{} 1}^H+ k_{1 \xrightarrow[]{} 0}^C k_{2 \xrightarrow[]{} 1}^H$ is a normalization factor.
The heat current, e.g. from the cold bath is given by
\bea
j_C= \left(p_0k_{0\to 1}^{C}-p_1k_{1\to 0}^{C} \right)(\epsilon_1-\epsilon_0),
\eea
and it is zero when plugging in the steady state populations.
From this simple analysis we conclude that in the ladder model
heat transport at weak coupling is due to processes beyond second order in the system-bath coupling.


\begin{thebibliography}{999}
\bibitem{comment2}
To the lowest order captured by the Redfield equation \cite{SegalQME}, 
the heat current is linear in the spectral density function $J(\omega)$.
Since this function is quadratic in the system-bath coupling energy $\lambda$,
the Redfield-level heat current is quadratic in $\lambda$.
\bibitem{comment3}
A measure for departing from second order perturbation method, which was
based on the reorganization energy
was suggested in Ref. \cite{Camille}, $\lambda^2/\Omega > T/10$. In our parameters,
this translates here to  $\lambda>1$.
However, since
a second order perturbative expansion totally fails for the ladder model,
it is not obvious that the reorganization criteria is the most relevant here.
\end{thebibliography}
\end{document}